\documentclass[superscriptaddress,twocolumn,showpacs,
amssymb,amsmath,nobibnotes,aps,prd,
nofootinbib]{revtex4-1}
\pdfoutput=1
\usepackage{graphicx,subfig,bm,caption,color,psfrag,hyperref}
\usepackage{amsfonts}
\usepackage{lipsum}
\usepackage[justification=justified, singlelinecheck=false]{caption}
\usepackage{mathtools}
\usepackage{verbatim}
\usepackage[normalem]{ulem}
\usepackage[dvipsnames]{xcolor}

\hypersetup{colorlinks,linkcolor={blue},citecolor={red},urlcolor={greenW}}  
\definecolor{greenW}{rgb}{0.0, 0.55, 0.1}

\begin{document}
%%%%%%%%%%%%%%%%%%%%%%%%%%%%%%%%%%%%%%%%%%%%%%%%%%%%%%%%%%%%%%%%%%%%%%%%%%
%%%%%%%%%%%%%%%%%%%%%%%%%%%%%%%%%%%%%%%%%%%%%%%%%%%%%%%%%%%%%%%%%%%%%%%%%%

\title{Shape of Dark Energy: Constraining Its Evolution with a General Parametrization}

\author{Dong Ha Lee}
\email{dhlee1@sheffield.ac.uk}
\affiliation{School of Mathematical and Physical Sciences, University of Sheffield, Hounsfield Road, Sheffield S3 7RH, United Kingdom}

\author{Weiqiang Yang}
\email{d11102004@163.com}
\affiliation{Department of Physics, Liaoning Normal University, Dalian, 116029, People's Republic of China}

\author{Eleonora Di Valentino}
\email{e.divalentino@sheffield.ac.uk}
\affiliation{School of Mathematical and Physical Sciences, University of Sheffield, Hounsfield Road, Sheffield S3 7RH, United Kingdom}

\author{Supriya Pan}
\email{supriya.maths@presiuniv.ac.in}
\affiliation{Department of Mathematics, Presidency University, 86/1 College Street, Kolkata 700073, India}
\affiliation{Institute of Systems Science, 
Durban University of Technology, Durban 4000, Republic of South Africa}

\author{Carsten van de Bruck}
\email{c.vandebruck@sheffield.ac.uk}
\affiliation{School of Mathematical and Physical Sciences, University of Sheffield, Hounsfield Road, Sheffield S3 7RH, United Kingdom}

%%%%%%%%%%%%%%%%%%%%%%%%%%%%%%%%%%%%%%%%%%%%%%%%%%%%%%%%%%%%%%%%%%%%
\begin{abstract}

We consider a general dark energy (DE) model parametrized by its equation-of-state (EoS), featuring three free parameters: $w_0$ (the present-day value of the DE EoS), $w_{\beta}$ (quantifying the dynamical nature of the DE EoS), and $\beta$ (governing various dynamical forms of the DE EoS). The key controlling parameter $\beta$ can recover several existing DE models in the literature, such as the Chevallier–Polarski–Linder (CPL) parametrization ($\beta = 1$), the logarithmic parametrization (in the limit $\beta \rightarrow 0$), and the linear parametrization ($\beta = -1$),  alongside generate a class of new DE parametrizations for other values of $\beta$. 
The resulting DE scenario is constrained using a suite of the latest cosmological probes, including Cosmic Microwave Background (CMB) temperature and polarization anisotropies from three different experiments (Planck 2018 and Atacama Cosmology Telescope combined with WMAP), CMB lensing, Baryon Acoustic Oscillations from DESI Year 2, and PantheonPlus from Type Ia supernovae. Our analyses reveal that stringent constraints on the DE parameters are obtained only when all cosmological probes are combined; otherwise, some parameters remain unconstrained. The present-day value of the DE EoS remains in the quintessence regime according to our results, and no significant evidence for a dynamical DE EoS is found. However, based on the $\Delta \chi^2$ and Bayesian evidence analyses, we observe a mild preference for the present three-parameter DE parametrization over the CPL parametrization when all cosmological probes are taken into account. Nonetheless, the Bayesian evidence difference remains below the threshold for statistical significance according to the revised Jeffreys scale, indicating that both models are effectively equally preferred by the data.

\end{abstract}

%%%%%%%%%%%%%%%%%%%%%%%%%%%%%%%%%%%%%%%%%%%%%%%%%%%%%%%%%%%%%%%%%%
\maketitle
%%%%%%%%%%%%%%%%%%%%%%%%%%%%%%%%%%%%%%%%%%%%%%%%%%%%%%%%%%%%%%%%%%

\section{Introduction}

One of the most significant discoveries in modern cosmology and astrophysics is the late-time accelerating expansion of the universe~\cite{SupernovaSearchTeam:1998fmf,SupernovaCosmologyProject:1998vns}. The intrinsic mechanism behind this accelerating expansion is not yet clear to the scientific community, and as a consequence, a large number of  cosmological hypotheses have been suggested. Usually, two common approaches are considered for describing this accelerating expansion: dark energy (DE)~\cite{Peebles:2002gy,Copeland:2006wr,Sahni:2006pa,Frieman:2008sn,Li:2011sd,Bamba:2012cp} and modified gravity (MG) theories~\cite{Nojiri:2006ri,Sotiriou:2008rp,DeFelice:2010aj,Capozziello:2011et,Clifton:2011jh,Koyama:2015vza,Cai:2015emx,Nojiri:2017ncd,Bahamonde:2021gfp}. The former approach is relatively simple, where some hypothetical energy component, such as a fluid or a scalar field, is introduced in the gravitational equations described by General Relativity (GR), while the latter approach relies on the fact that GR does not hold on all scales, and hence modifications of GR leading to new geometrical terms are considered for a possible explanation of the accelerated expansion of the universe. These mechanisms have attracted considerable attention among cosmologists and a large number of models have been proposed (for an incomplete list of references, see for instance ~\cite{Bilic:2001cg,Padmanabhan:2002sh,Capozziello:2002rd,Bagla:2002yn,Gorini:2002kf,Dolgov:2003px,Carroll:2003st,Farrar:2003uw,Nojiri:2003ft,Carroll:2004de,Li:2004rb,Guo:2004fq,Nojiri:2005vv,Pavon:2005yx,Wei:2005nw,Gumjudpai:2005ry,Elizalde:2005ju,Brevik:2005bj,Cognola:2006eg,Ferraro:2006jd,Li:2007jm,Nojiri:2007cq,Wei:2007ty,DeFelice:2008wz,Feng:2009jr,Saridakis:2009bv,Dent:2010nbw,Harko:2011kv,Olmo:2011uz,Gubitosi:2012hu,Bloomfield:2012ff,Maggiore:2013mea,Yang:2014gza,Maggiore:2014sia,Kamionkowski:2014zda,Sola:2015rra,Astashenok:2015haa,Katsuragawa:2016yir,Lin:2017oow,Heisenberg:2018yae,Langlois:2018dxi,DiValentino:2019ffd,Braglia:2020auw,Motta:2021hvl,Fernandes:2022zrq,Raveri:2021dbu,Kumar:2023bqj,Heisenberg:2023lru,Chudaykin:2024gol,Yang:2025ume,Gialamas:2025pwv,Mishra:2025goj,Wolf:2024stt,Dhawan:2025mer}).

Among this list of models, the $\Lambda$-Cold Dark Matter ($\Lambda$CDM) cosmological model constructed within the framework of GR agrees well with a series of astronomical datasets. However, the $\Lambda$CDM model faces several theoretical and observational challenges. Starting from the cosmological constant problem~\cite{Weinberg:1988cp}, the cosmic coincidence problem~\cite{Zlatev:1998tr}, to the Hubble constant tension~\cite{Verde:2019ivm,DiValentino:2020zio,DiValentino:2021izs,Perivolaropoulos:2021jda,Schoneberg:2021qvd,Shah:2021onj,Abdalla:2022yfr,DiValentino:2022fjm,Kamionkowski:2022pkx,Giare:2023xoc,Hu:2023jqc,Verde:2023lmm,DiValentino:2024yew,Perivolaropoulos:2024yxv,CosmoVerse:2025txj}, the $\Lambda$CDM model has been criticized at different levels and suggestions for revising this cosmological model have been proposed.

In this article, we adopt the validity of GR and focus on DE models. In the context of GR, a possible revision of the $\Lambda$CDM cosmology is performed through the modification of the DE sector which can be performed through a number of different methods. Here, we choose to modify the DE sector in terms of its barotropic equation-of-state (EoS) $w_{\rm DE} = p_{\rm DE}/\rho_{\rm DE}$, where $p_{\rm DE}$, $\rho_{\rm DE}$ are respectively the pressure and energy density of the DE fluid.

In the $\Lambda$CDM paradigm, DE, i.e. $\Lambda$, corresponds to the vacuum energy modeled by a homogeneous fluid with a constant barotropic EoS $w_{\rm DE} = -1$. The simplest modification of $\Lambda$CDM is to assume a DE model in which $w_{\rm DE}$ is either a constant not equal to $-1$, or time-varying. The latter proposal is very attractive for several reasons. From the phenomenological point of view, dynamical $w_{\rm DE}$ offers more freedom to test the nature of DE. With the ever increasing sensitivity of the astronomical datasets to cosmological models, it appears to be sensible to allow the observational data to determine the fate of $w_{\rm DE}$ -- whether it be constant or dynamical. As a result, there has been significant work within the community to work with various phenomenologically motivated models for $w_{\rm DE}$~\cite{Cooray:1999da,Efstathiou:1999tm,Chevallier:2000qy,Linder:2002et,Wetterich:2004pv,Feng:2004ff,Hannestad:2004cb,Xia:2004rw,Gong:2005de,Jassal:2005qc,Nesseris:2005ur,Liu:2008vy,Barboza:2008rh,Barboza:2009ks,Ma:2011nc,Sendra:2011pt,Feng:2011zzo,Barboza:2011gd,DeFelice:2012vd,Feng:2012gf,Wei:2013jya,Magana:2014voa,Akarsu:2015yea,Pan:2016jli,DiValentino:2016hlg,Nunes:2016plz,Nunes:2016drj,Magana:2017usz,Yang:2017alx,Pan:2017zoh,Panotopoulos:2018sso,Yang:2018qmz,Jaime:2018ftn,Das:2017gjj,Yang:2018prh,Li:2019yem,Yang:2019jwn,Pan:2019hac,Tamayo:2019gqj,Pan:2019brc,DiValentino:2020naf,Rezaei:2020mrj,Perkovic:2020mph,Banihashemi:2020wtb,Jaber-Bravo:2019nrk,Benaoum:2020qsi,Yang:2021eud,Jaber:2021hho,Alestas:2021luu,Yang:2022klj,Escudero:2022rbq,Castillo-Santos:2022yoi,Yang:2022kho,Dahmani:2023bsb,Escamilla:2023oce,Rezaei:2023xkj,Adil:2023exv,LozanoTorres:2024tnt,Singh:2023ryd,Rezaei:2024vtg,Reyhani:2024cnr,DESI:2024mwx,Cortes:2024lgw,Shlivko:2024llw,Luongo:2024fww,Yin:2024hba,Gialamas:2024lyw,Dinda:2024kjf,Najafi:2024qzm,Wang:2024dka,Ye:2024ywg,Tada:2024znt,Carloni:2024zpl,Chan-GyungPark:2024mlx,DESI:2024kob,Ramadan:2024kmn,Notari:2024rti,Orchard:2024bve,Hernandez-Almada:2024ost,Pourojaghi:2024tmw,Giare:2024gpk,Reboucas:2024smm,Giare:2024ocw,Chan-GyungPark:2024brx,Menci:2024hop,Li:2024qus,Li:2024hrv,Notari:2024zmi,Gao:2024ily,Fikri:2024klc,Jiang:2024xnu,Zheng:2024qzi,Gomez-Valent:2024ejh,RoyChoudhury:2024wri,Lewis:2024cqj,Paliathanasis:2025cuc,Wolf:2025jlc,Shajib:2025tpd,Giare:2025pzu,Chaussidon:2025npr,Kessler:2025kju,Pang:2025lvh,RoyChoudhury:2025dhe,Scherer:2025esj,Specogna:2025guo,Cheng:2025lod,Cheng:2025hug,Ozulker:2025ehg}.

Recently, there has been a renewed interest in investigating the form of $w_{\rm DE}$ after the recent data releases of the Dark Energy Spectroscopic Instrument (DESI)~\cite{DESI:2024mwx,DESI:2025zgx}, which appeared to indicate some preference of dynamical $w_{\rm DE}$~\cite{DESI:2024mwx,DESI:2025zgx,Giare:2024gpk,Giare:2025pzu}. Although the DESI analysis assumed the Chevallier-Polarski-Linder parametrization (CPL) for $w_{\rm DE}$, the evidence of dynamical $w_{\rm DE}$ was found across a variety of well-known parametrizations~\cite{Giare:2024gpk,Wolf:2025jlc, luPreferenceEvolvingDark2025}. Therefore, the data appear to prefer a dynamical DE, over a constant, regardless of the parameterization used.

Furthermore, even before the DESI data release, non-parametric approaches also indicated the evidence for the time-dependent nature of $w_{\rm DE}$~\cite{Zhao:2017cud,Zhang:2019jsu,Escamilla:2024fzq} (also see~\cite{Ormondroyd:2025exu,Ormondroyd:2025iaf}, that appeared after the DESI release). As the non-parametric approaches are unable to extract the exact functional form for $w_{\rm DE}$, parametrizing $w_{\rm DE}$ has been a preferred route to study the evolution of DE at the background and perturbative levels.

In the present article, we consider a generalized DE EoS with three free parameters proposed by Barboza et al.~\cite{Barboza:2009ks} that quickly retrieves three well-known and most-used DE parametrizations in the literature, namely CPL parametrization~\cite{Chevallier:2000qy,Linder:2002et}, linear parametrization~\cite{Cooray:1999da}, and the logarithmic parametrization~\cite{Efstathiou:1999tm}. According to the existing records in the literature, this particular DE model has not received much attention~\cite{Ma:2017xlk,Yang:2017yme} and how this model behaves at the perturbative level has mostly been ignored. 

However, we have constrained the generalized DE scenario using several latest observational datasets, including two versions of the cosmic microwave background (CMB) data, one from Planck 2018, and one from ACT DR6 in combination with the WMAP measurements; CMB lensing data from the ACT DR6 lensing likelihoods, in combination with the Planck PR4 lensing data; baryon acoustic oscillations (BAO) from DESI DR2; and Pantheon+ from Type Ia supernovae. The results of our analyses indicate that there is enough room to work in the dark sector using the parametrized framework.

The structure of this work is as follows. In Section~\ref{sec-setup}, we describe the basic cosmological equations, propose the model, and present its evolution at the background and perturbation levels. In Section~\ref{sec:data}, we present the astronomical datasets used to constrain the DE model. Section~\ref{sec-results} describes the results of our analysis. Finally, with Section~\ref{sec-summary}, we close the article by presenting the main findings.

\begin{figure*}
    \centering
    \subfloat[Effect of varying $\beta$ over the full range of priors on the equation of state $w_{\rm DE}(z)$ for fixed $w_\beta = \pm0.5$ and $w_0 = -1$.\label{fig:vary_beta}]{%
    \includegraphics[width=0.49\textwidth]{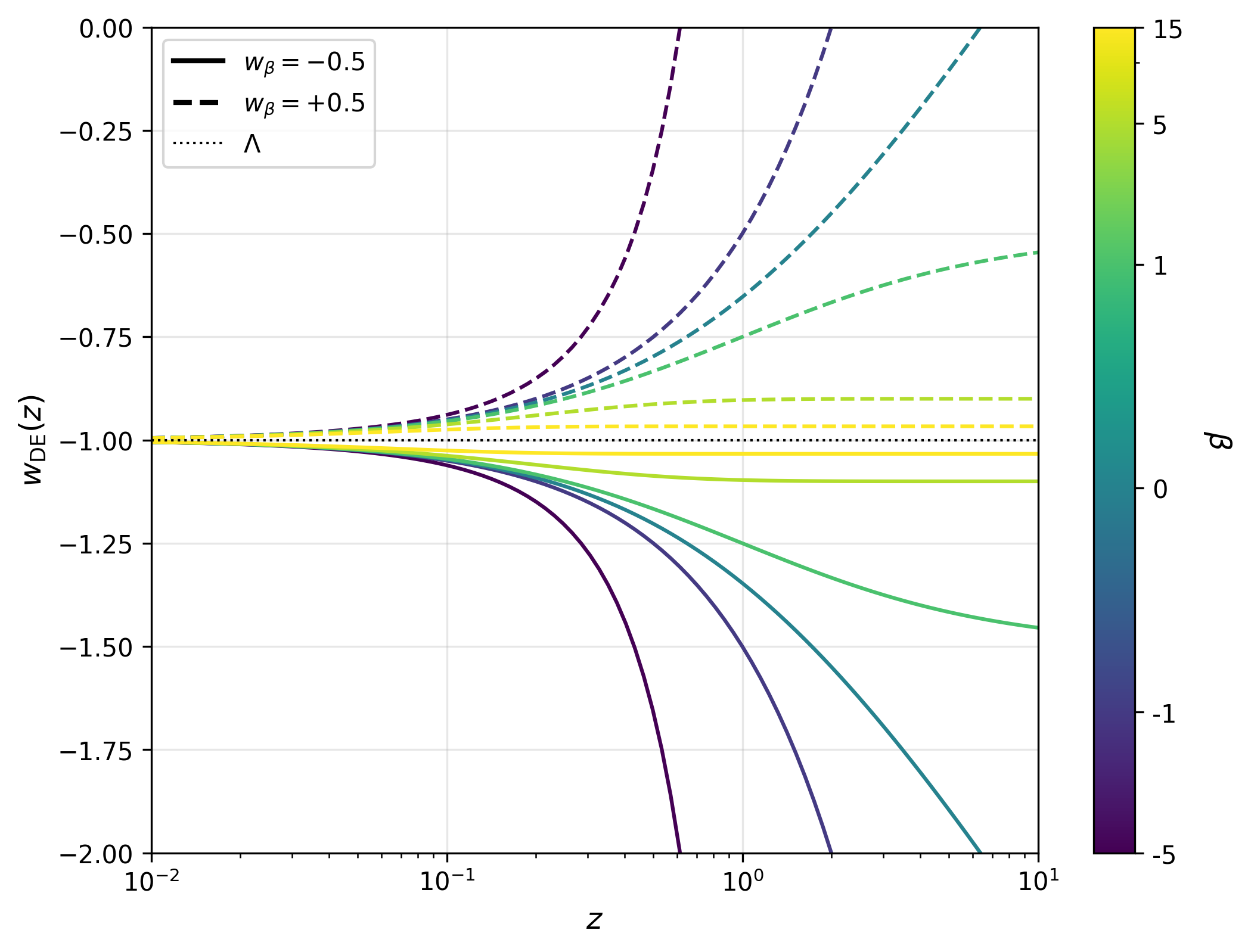}
    }
    \hfill
    \subfloat[Effect of varying $w_\beta$ over the full range of priors on the equation of state $w_{\rm DE}(z)$ for fixed $\beta = \pm1$ and $w_0 = -1$.\label{fig:vary_wb}]{%
        \includegraphics[width=0.49\textwidth]{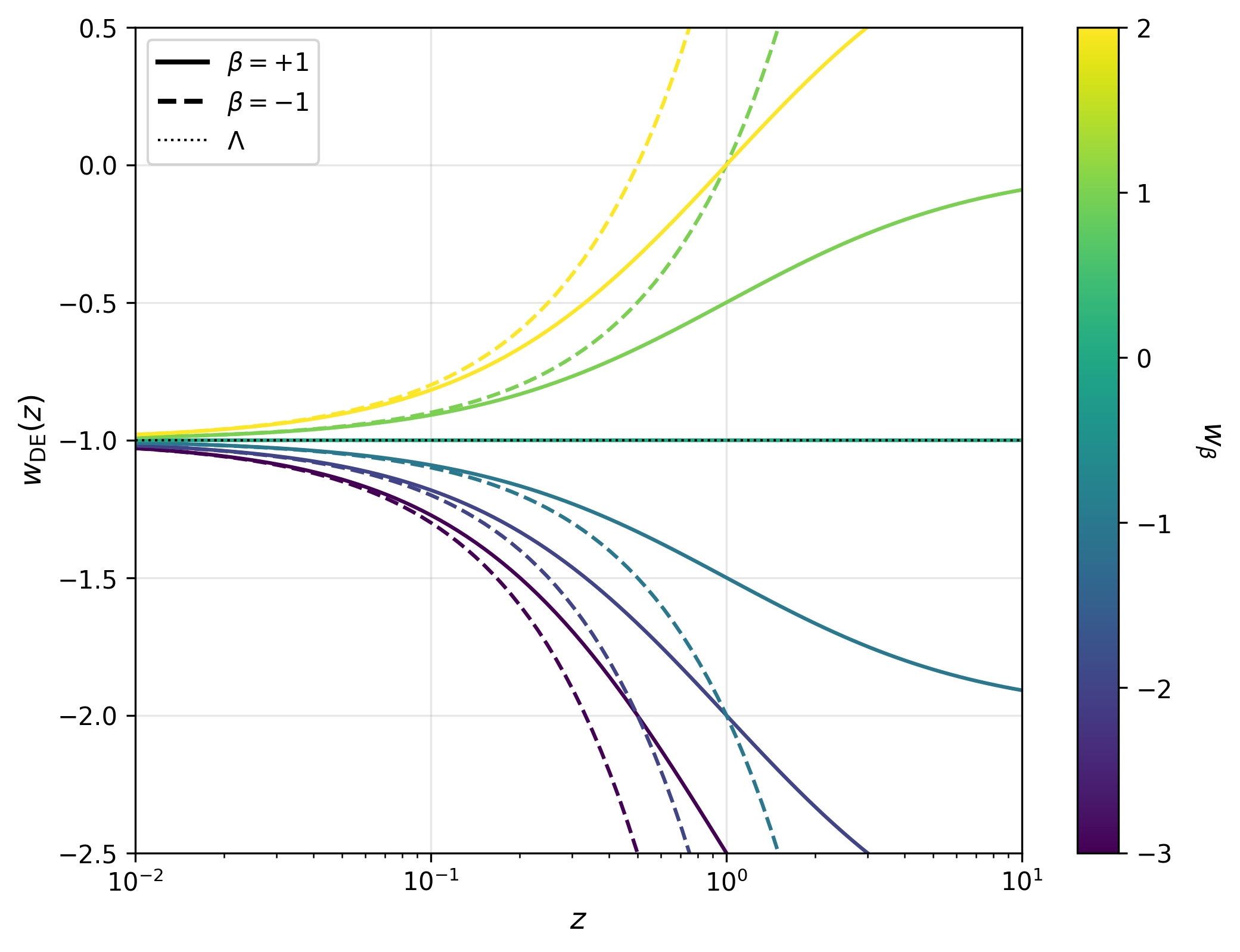}
    }
    \caption{The effects of $\beta$  and $w_\beta$ on $w_{\rm DE}(z)$ when varied across a wide range on those parameters.}
    \label{fig:DE_eos}
\end{figure*}
\begin{figure*}
    \centering
    \subfloat[Effect of varying $\beta$ on the CMB $TT$ power spectrum.\label{fig:tt}]{%
        \includegraphics[height=6.5cm]{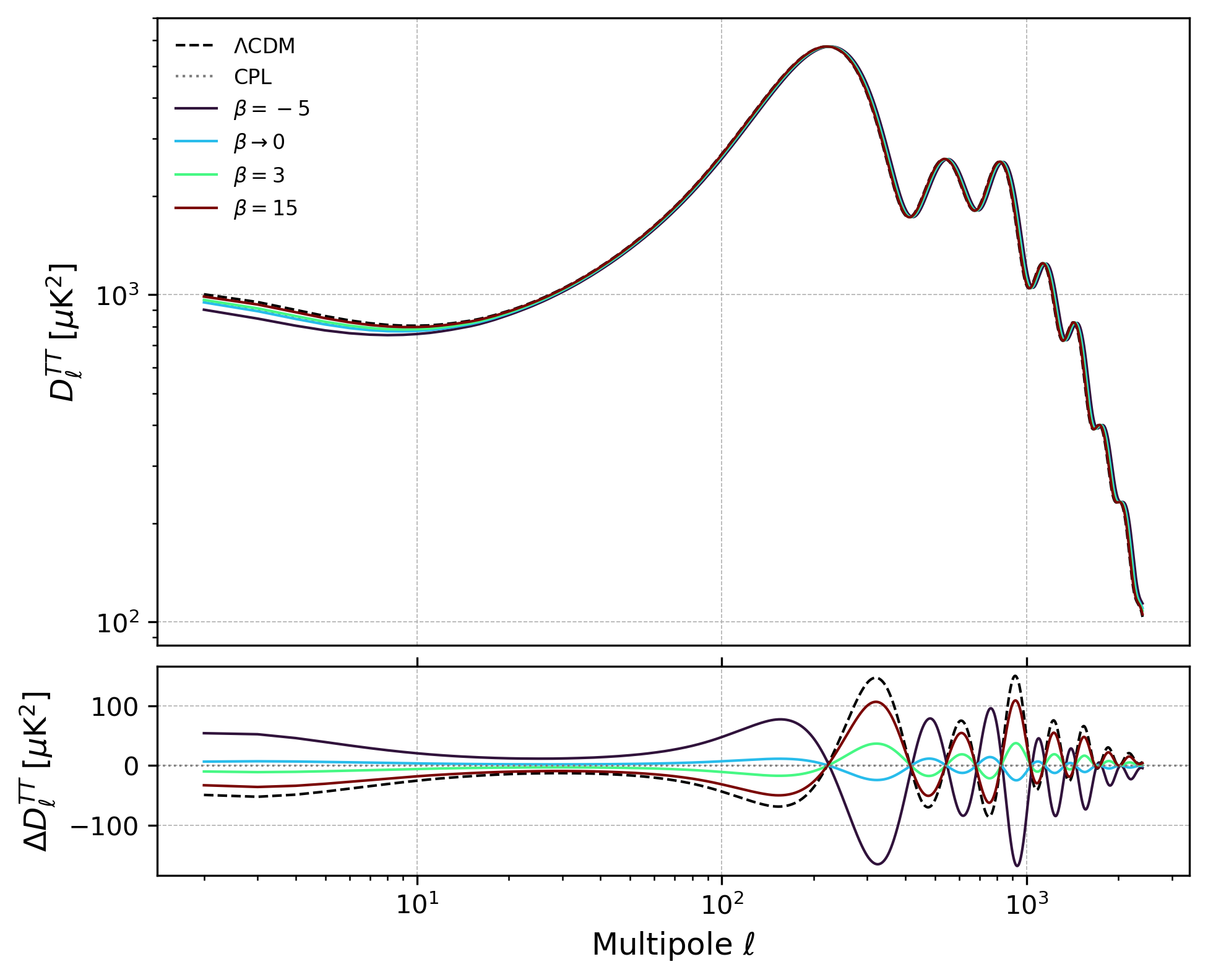}
    }
    \hfill
    \subfloat[Effect of varying $\beta$ on the CMB $EE$ power spectrum.\label{fig:ee}]{%
    \includegraphics[height=6.5cm]{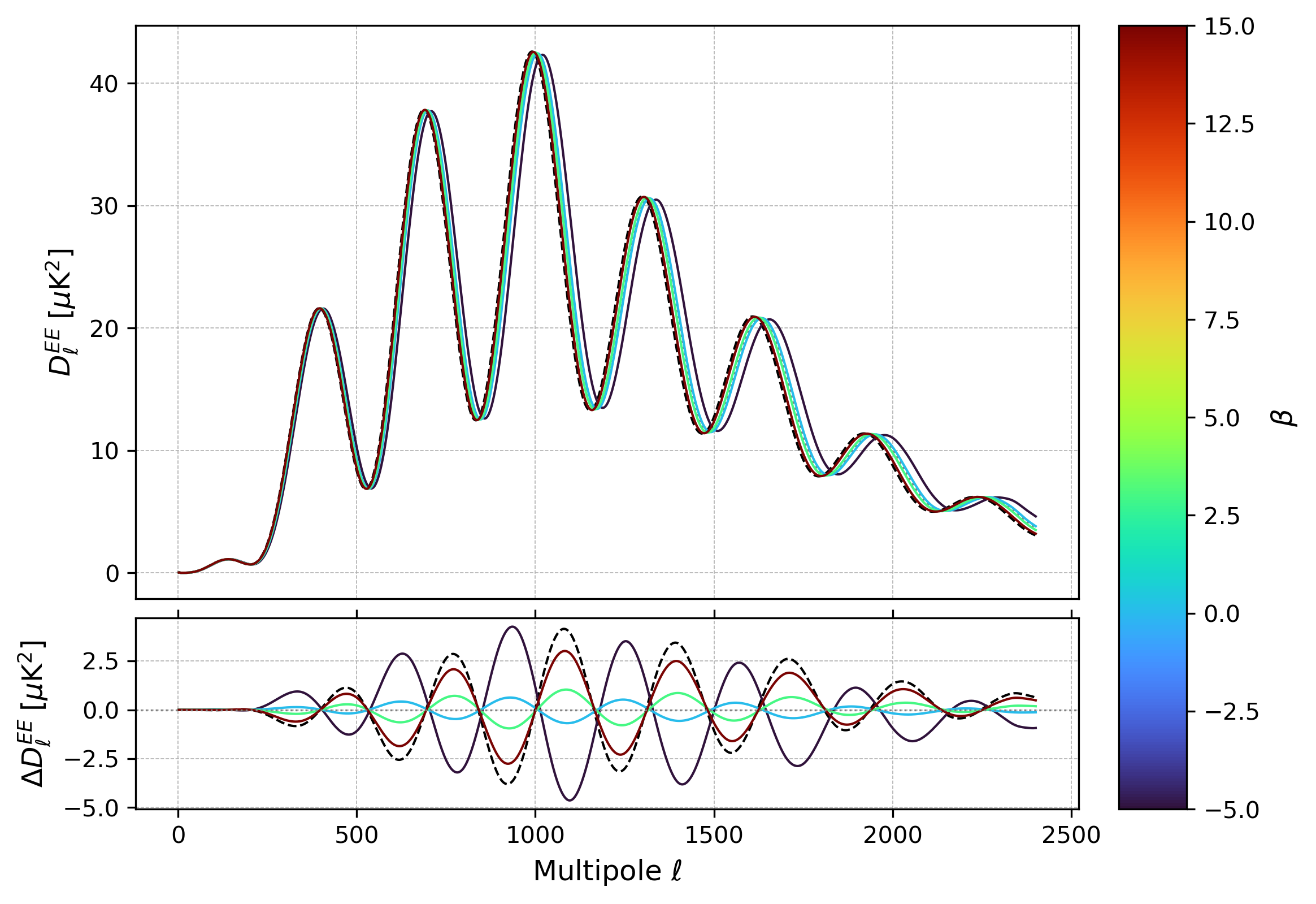}
    }
    \caption{Effects of the present DE parametrization, alongside the $\Lambda$CDM and CPL models, on the CMB TT (left) and CMB EE (right) power spectra for different values of $\beta$. The plots are obtained by fixing the $\Lambda$CDM parameters to their best-fit values from fitting ACT DR6 + Planck primary CMB data, combined with CMB lensing from ACT DR6 and Planck PR4 (NPIPE). The dark energy parameters are fixed to $w_0 = -1$ and $w_{\beta} = -0.8$, allowing only $\beta$ to vary. The residual plots are shown with respect to the CPL model ($\beta=1$) as the reference.}
    \label{fig:CMB-TT-EE}
\end{figure*}

\section{Dark energy: set-up}
\label{sec-setup}

As mentioned in the introduction, we assume the validity of GR to describe the gravitational sector of the universe and consider the spatially flat Friedmann-Lema\^itre-Robertson-Walker line element for the geometrical description of the universe, which takes the form 
\begin{equation}
ds^2 = - dt^2 + a^2(t) (dx^2 + dy^2 + dz^2),
\end{equation}
where $t$ is cosmic time and $a(t)$ is the scale factor, describing the evolution of the universe. Assuming further that the matter distribution of the universe is minimally coupled to GR, and that there is no interaction between any two fluids in this distribution, the dynamics of the universe can be described by the Friedmann equations:
\begin{eqnarray}
H^2 &=& \frac{8\pi G}{3} (\rho_r + \rho_b + \rho_{\nu} + \rho_{\rm c} + \rho_{\rm DE}), \\
2 \dot{H} + 3 H^2 &=& -8 \pi G (p_r + p_b + p_{\nu} + p_{\rm c} + p_{\rm DE}),
\end{eqnarray}
where $H = \dot a/a$, a dot denotes derivative with respect to cosmic time and $\rho_r$, $\rho_b$, $\rho_{\nu}$, $\rho_{\rm c}$, and $\rho_{\rm DE}$ are respectively the energy densities of radiation, baryons, neutrinos, pressureless dark matter (alternatively known as cold dark matter), and dark energy, while $p_r$, $p_b$, $p_{\nu}$, $p_{\rm c}$, and $p_{\rm DE}$ are the respective pressure terms (note that $p_{\rm c} = 0$). All fluids are barotropic, for neutrinos the sum of neutrino masses, $\sum m_\nu$, is fixed to $0.06$~eV, and the number of neutrino species $N_{\textrm{eff}}$ is fixed to $3.044$. The conservation equation of each individual fluid in the absence of interaction gives
\begin{equation}
\dot{\rho}_i + 3 H (\rho_i + p_i) = 0,
\end{equation}
where $i$ runs over all components. Using the conservation equation for DE, its energy density evolution can be found as 
 
\begin{eqnarray}\label{evol-eqn-DE}
\frac{\rho_{\rm DE}}{\rho_{\rm DE,0}} =  a^{-3} \, \exp\left(-3 \int_{1}^{a}\frac{w_{\rm DE} (a')}{a'}\,  da'\right),
\end{eqnarray}
where $\rho_{\rm DE,0}$ is the present value of $\rho_{\rm DE}$. Thus, once $w_{\rm DE}$ is known, the evolution of $\rho_{\rm DE}$ is automatically determined. 
Here we consider a general parameterization for $w_{\rm DE}$ as proposed in~\cite{Barboza:2009ks}:
\begin{equation}\label{general-eos}
w_{\rm DE} (a) = w_0 - w_{\beta} \left[\frac{a^{\beta}-1}{\beta}\right],
\end{equation}
where $w_0$ is the current value of $w_{\rm DE}$, $\beta$ and $w_{\beta}$ are real constants.
\par

We note that this parametrization is divergent for $\beta=0$, but in the limit $\beta \rightarrow 0$ it tends to the logarithmic parametrization $w_{\rm DE}(a) = w_0 - w_{\beta} \ln(a)$~\cite{Efstathiou:1999tm}. This three-parameter model for the DE equation of state was of particular interest because it is able to reproduce a set of familiar two-parameter parametrizations by fixing one of the parameters.\footnote{There are many ways to construct several two-parameter DE parametrizations. One of the possible approaches is to deal with the two parameters $(\beta, w_{\beta})$ while fixing $w_0 = -1$. This set of parametrizations measures the deviation from the cosmological constant allowed by the recent observational data.} The hope of this model was that, by allowing the three parameters to vary, the cosmological datasets would be able to ``select'' a preferred model of DE based on its fit to the data.
\par
For example, the logarithmic parametrization, $w_{\rm DE}(a) = w_0 - w_{\beta} \ln(a)$~\cite{Efstathiou:1999tm}, discussed above for $\beta \rightarrow 0$; the CPL parametrization, $w_{\rm DE} = w_0 + w_{\beta}(1-a)$~\cite{Chevallier:2000qy,Linder:2002et}, for $\beta = 1$; and, for $\beta = -1$, the linear parametrization in terms of the cosmological redshift, $w_{\rm DE} = w_0 + w_{\beta} z$~\cite{Cooray:1999da}.
\par
Consequently, each point in the $w_\beta - \beta$ plane corresponds to a different evolution of the equation of state, with $w_0$ shifting the present-day value of the equation of state. This is more clearly shown in Fig.~\ref{fig:DE_eos}, where on the left we see $w_{\rm DE}(z)$ versus $z$ for varying $\beta$ over a wide range of values. The horizontal dotted line corresponds to $w_{\rm DE} = -1$, that is, $\Lambda$, and the curves in the upper and lower halves correspond to positive and negative values of $w_\beta$, respectively. It must be noted that here we have fixed $w_\beta = \pm 0.5$ and $w_0 = -1$ for visualizing a subset of possible models produced by Eq.~(\ref{general-eos}) and how the choice of $\beta$ can affect the dynamics of the equation of state.
\par
The right figure in Fig.~\ref{fig:DE_eos} displays the effect of varying $w_\beta$ for fixed values of $\beta = \pm 1$. The two figures indicate that $\beta$ fixes the rate of the evolution of DE, while $w_\beta$ determines the direction of the evolution.
\par
Now, for this model, the DE evolution becomes
\begin{eqnarray}\label{rho-DE}
\frac{\rho_{\rm DE}}{\rho_{\rm DE,0}} =  a^{-3 \left(1+w_0 + \frac{w_{\beta}}{\beta} \right)}\, \exp \left[ \frac{3 w_{\beta}}{\beta^2}  \left(a^{\beta} - 1 \right) \right].
\end{eqnarray}
Therefore, considering the evolution of DE in Eq.~(\ref{rho-DE}) together with the evolution of the remaining fluids present in the Friedmann equation, one can find the evolution of the expansion history of the universe at the background level. 
To study the formation of structures in the universe, we must examine the evolution of cosmological perturbations in this model. In what follows, we present the key equations at the perturbative level.

We consider the following line element  
in the synchronous gauge~\cite{Ma:1995ey}, using here conformal time instead of cosmic time:
\begin{equation}\label{eq12}
ds^2 = a^2(\tau) \left[-d\tau^2 + (\delta_{ij} + h_{ij}) dx^i dx^j \right],
\end{equation}
where $\delta_{ij}$ and $h_{ij}$ are the unperturbed and perturbed spatial parts of the metric tensor, respectively, and $\tau$ denotes conformal time. 
The density contrast of the $i$-th fluid is defined as $\delta_i = \delta \rho_i / \rho_i$, and $\theta_i = i \kappa^j v_j$ represents the divergence of the fluid’s velocity. From energy momentum conservation, one can derive the following set of equations~\cite{Ma:1995ey}:
\begin{eqnarray}
\delta'_{i} & = & - (1+ w_{i})\, \left(\theta_{i}+ \frac{h'}{2}\right) - 
3\mathcal{H}\left(\frac{\delta P_i}{\delta \rho_i} - w_{i} \right)\delta_i \nonumber \\ 
& & -  9 \mathcal{H}^2\left(\frac{\delta P_i}{\delta \rho_i} - c^2_{a,i} \right) (1+w_i) 
\frac{\theta_i}
{{\kappa}^2}, \label{per1} \\
\theta'_{i} & = & - \mathcal{H} \left(1- 3 \frac{\delta P_i}{\delta
\rho_i}\right)\theta_{i} 
+ \frac{\delta P_i/\delta \rho_i}{1+w_{i}}\, {\kappa}^2\, \delta_{i} 
-{\kappa}^2\sigma_i,\label{per2}
\end{eqnarray}
where a prime denotes differentiation with respect to conformal time, and $\mathcal{H} \equiv a'/a$ is the conformal Hubble parameter. The quantities $\kappa$, $h$, and $\sigma_i$ represent the comoving wavenumber, the synchronous gauge metric perturbation, and the anisotropic stress of the $i$-th fluid, respectively.\footnote{In this article we set $\sigma_i = 0$, i.e., we assume no anisotropic stress.}
The quantity $\delta P_i / \delta \rho_i$ denotes the square of the sound speed in the fluid’s rest frame. For DE, this is given by $c^2_{\rm s,DE}$ (i.e., $\delta P_{\rm DE} / \delta \rho_{\rm DE} = c^2_{\rm s,DE}$). The adiabatic sound speed of the $i$-th fluid is defined as
\begin{equation}
c^2_{a,i} = w_i - \frac{w_i^{\prime}}{3 \mathcal{H} (1+w_i)}.
\end{equation}
In this work, we fix $c^2_{\rm s,DE} = 1$, as is typical in models with minimally coupled scalar fields.

We close this section with Fig.~\ref{fig:CMB-TT-EE}, which shows the CMB TT (left plot of Fig.~\ref{fig:CMB-TT-EE})
and CMB EE (right plot of Fig.~\ref{fig:CMB-TT-EE})
power spectra for different values of $\beta$ in the present DE parametrization. Additionally, the lower panels on each side (left and right) show the deviations with respect to the CPL model ($\beta=1$). 
As depicted in Fig.~\ref{fig:CMB-TT-EE}, we observe that as $\beta$ increases in either the positive or negative direction, noticeable effects on the CMB TT and EE spectra emerge. Although such effects may not be clearly evident from the upper panels of the left and right plots in Fig.~\ref{fig:CMB-TT-EE}, they become much more apparent in the respective residual plots. 
These imprints on the CMB spectra clearly indicate that the present DE parametrization can influence the CMB and, therefore, may play a significant role in the evolution of the universe.

\begin{table}[!ht]
	\begin{center}
		\footnotesize
		\renewcommand{\arraystretch}{1.5}
		\begin{tabular}{l@{\hspace{0. cm}}@{\hspace{2 cm}} c}
			\hline\hline
			\textbf{Parameter} & \textbf{Prior} \\
			\hline\hline
			$\Omega_{\rm b} h^2$ & $[0.005\,,\,0.1]$ \\
			$\Omega_{\rm c} h^2$ & $[0.005\,,\,0.99]$ \\
			$\tau_{\rm reio}$ & $[0.01, 0.8]$ \\
			$100\,\theta_s$ & $[0.5\,,\,10]$ \\
			$\log(10^{10}A_{\rm S})$ & $[1.61\,,\,3.91]$ \\
			$n_{\rm s}$ & $[0.8\,,\, 1.2]$ \\
			$w_{0}$ & $[-3\,,\, 1]$ \\
            $w_{\beta}$ & $[-3\,,\,2]$ \\
            $\beta$ & $[-5\,,\, 15]$ \\
            
			\hline\hline
		\end{tabular}
		\caption{Flat priors on the model parameters used during the statistical analysis.}
		\label{table:priors}
	\end{center}
\end{table}

\section{The Observational Data}
\label{sec:data}

In this section, we summarize the cosmological datasets used for the statistical simulations. 

\begin{enumerate}

\item We use the Planck 2018 primary CMB measurements (Planck Commander low-$\ell$ TT, Planck SimAll low-$\ell$ EE, and Planck Plik high-$\ell$ TT, TE, EE likelihoods), which are labeled as ``P'' in the results~\cite{Planck:2019nip}.

\par We also include the primary CMB measurements from ACT DR6 (ACT-lite likelihoods)~\cite{ACT:2025fju}, combined with the WMAP data truncated at $\ell > 23$, in order to obtain a Planck-independent primary CMB dataset. This combination is labeled as ``AW'' in the results~\cite{WMAP:2012fli}. Due to ACT's lacking of constraining power at low multipoles in the EE spectrum, we impose a Gaussian prior on the optical depth to reionization, $\tau_{\rm reio} = 0.0566 \pm 0.0058$, derived from the Planck Sroll2 low-$\ell$ EE likelihood~\cite{Pagano:2019tci}.

\item In addition to the primary CMB data, we incorporate the CMB lensing likelihood from the ACT DR6 lensing release~\cite{ACT:2023kun, ACT:2023dou}, combined with the Planck PR4 NPIPE lensing likelihood~\cite{Carron:2022eyg}, in all our analyses. This dataset is labeled as ``L'' in the results.

\item We use the latest measurements of BAO features imprinted on large-scale structures by the DESI collaboration, as reported in their second data release (DR2)~\cite{DESI:2025zgx}. This dataset, labeled as ``B'' in the results, includes a comprehensive set of 16 BAO measurements of $D_M/r_{\rm d}$ and $D_H/r_{\rm d}$ across the redshift range $0.4 < z < 4.2$, as well as one low-redshift measurement of $D_V/r_{\rm d}$ in the range $0.1 < z < 0.4$. These measurements are obtained using multiple tracers, including emission line galaxies (ELGs), luminous red galaxies (LRGs), quasars (QSOs), and Ly$\alpha$ forest data.

\item We use the Pantheon+ compilation~\cite{Brout:2022vxf} of Type Ia supernova (SNIa) data, labeled as ``S'' in our results. This dataset includes 1701 light curves corresponding to 1550 unique SNIa, spanning the redshift range $z \in [0.001, 2.26]$. It incorporates significant improvements over previous compilations, such as enhanced photometric calibration, a refined treatment of systematic uncertainties, and a consistent reanalysis of all contributing samples.

\end{enumerate}

For the statistical analysis, we modified the publicly available package \texttt{CAMB} (Code for Anisotropies in the Microwave Background)~\cite{Lewis:1999bs,Howlett:2012mh} and used another publicly available sampler, \texttt{Cobaya} (Code for Bayesian Analysis)~\cite{Torrado:2020dgo}, to perform the Markov Chain Monte Carlo (MCMC) analyses. The chains were considered converged when the Gelman–Rubin criterion~\cite{Gelman:1992zz} satisfied the threshold $R-1 < 0.01$.

The present DE parametrization involves nine free parameters. The first six are the standard $\Lambda$CDM parameters: $\Omega_{\rm b} h^2$ (the present-day physical baryon density), $\Omega_{\rm c} h^2$ (the present-day physical cold dark matter density), $\tau_{\rm reio}$ (the optical depth to reionization), $100\theta_{\mathrm{s}}$ (where $\theta_{\mathrm{s}}$ is the angular size of the sound horizon at recombination), $\log(10^{10} A_{\mathrm{s}})$ (with $A_{\mathrm{s}}$ the amplitude of primordial scalar perturbations), and $n_{\mathrm{s}}$ (the scalar spectral index). The remaining three parameters describe the dark energy sector: $w_0$, $w_\beta$, and $\beta$. 
In Table~\ref{table:priors}, we list the flat priors\footnote{For dataset combinations involving the AW CMB data, we instead imposed the Gaussian prior on $\tau_{\rm reio}$ described earlier.} imposed on the free parameters used in our analyses.\footnote{The choice of the priors, specifically on $\beta$ and $w_{\beta}$, is crucial for the numerical simulations. For the regions of $\beta$ and $w_{\beta}$ used in Fig.~\ref{fig:DE_eos}, since the evolution of $w_{\rm DE}(z)$ did not exhibit any diverging nature, they may serve as viable prior ranges for the observational analyses.  }

With the converged MCMC chains from \texttt{Cobaya}, we can perform model comparison tests between our DE parameterization and the CPL model to understand the effects of introducing an additional degree of freedom in the DE equation of state. We can perform a goodness of fit test of two models $i$ and $j$ to a given dataset by finding the minimum $\chi^2$ for each model and computing $\Delta \chi^2_{ij} = \chi^2_\mathrm{i} - \chi^2_\mathrm{j}$. A positive $\Delta\chi^2$ indicates that model $j$ is a better fit to the data than model $i$. If models $i$ and $j$ are nested with $k$ additional parameters in model $j$, we would expect $\Delta\chi^2_{ij}$ to follow a $\chi^2$ distribution with $k$ degrees of freedom.

Finally, we perform Bayesian model comparison by computing the logarithm of the Bayesian evidence, $\ln \mathcal{Z}$, using \texttt{MCEvidence}~\citep{Heavens:2017afc} via the \texttt{Cobaya} interface provided in the \texttt{wgcosmo} repository~\citep{giare2025wgcosmo}. For a given model $\mathcal{M}_i$ with parameter vector $\Theta$, the Bayesian evidence is
\begin{equation}
\mathcal{Z}_i = \int \mathcal{L}(D|\Theta, \mathcal{M}_i)\, \pi(\Theta|\mathcal{M}_i)\, \mathrm{d}\Theta,
\end{equation}
where $\mathcal{L}$ is the likelihood and $\pi$ the prior.
Model comparison is based on the Bayes factor $\mathcal{Z}_{ij} = \mathcal{Z}_i / \mathcal{Z}_j$, expressed as the relative log-evidence:
\begin{equation}
\Delta \ln \mathcal{Z}_{ij} = \ln \mathcal{Z}_i - \ln \mathcal{Z}_j.
\end{equation}
Positive values of $\Delta \ln \mathcal{Z}_{ij}$ favor model $i$ over model $j$.
We interpret $\Delta \ln \mathcal{Z}$ using the revised Jeffreys’ scale~\citep{Kass:1995loi}: values in $[0,1]$ are \textit{inconclusive}, $[1,2.5]$ indicate \textit{weak}, $[2.5,5]$ \textit{moderate}, $[5,10]$ \textit{strong}, and $>10$ \textit{very strong} evidence.

\begin{table*}%[h]
    \centering
    \begin{tabular}{cccccccccc}
    \hline
    Parameters&P-L& P-LS & P-LB & P-LBS\\
    \hline\hline
    $100\theta_\mathrm{MC}$ & $1.04100\pm 0.00031$ &$1.04091\pm 0.00031$&$1.04102\pm 0.00029$&$1.04108\pm 0.00029$ 
    \\
    $\Omega_\mathrm{b} h^2$ & $0.02244\pm 0.00014$ & $0.02238\pm 0.00015$ & $0.02243\pm 0.00014$ & $0.02246\pm 0.00014$
    \\
    $\Omega_\mathrm{c} h^2$&$0.1194\pm 0.0012$&$0.1201\pm 0.0012$&$0.11942\pm 0.00099$&$0.1189\pm 0.0010$
    \\
    $n_\mathrm{s}$&$0.9671\pm 0.0041$&$0.9654\pm 0.0042$&$0.9671\pm 0.0039$&$0.9683\pm 0.0039$
    \\
    $\tau_\mathrm{reio}$&$0.0526\pm 0.0073$&$0.0548\pm 0.0075$&$0.0551\pm 0.0075$&$0.0571\pm 0.0077$
    \\
    $\log(10^{10} A_\mathrm{s})$&$3.038\pm 0.013$&$3.046\pm 0.014$&$3.045\pm 0.014$&$3.049\pm 0.014$
    \\
    \hline
    $w_0$&$-1.20^{+0.25}_{-0.71}$&$-0.878\pm 0.092$&$-0.54^{+0.26}_{-0.21}$&$-0.842^{+0.060}_{-0.083}$
    \\
    $w_\beta$&$< -0.505$&$-1.0^{+1.1}_{-1.0}$&$< -1.16$&$-0.94^{+0.94}_{-0.31}$
    \\
    $\beta$&$< 7.83$&$> 4.80$&$2.0^{+1.2}_{-2.6}$&$2.8^{+2.8}_{-5.9}$
    \\
    \hline
    $H_0$&$> 79.3$&$67.27^{+0.88}_{-1.2}$&$65.0^{+1.8}_{-2.5}$&$67.71\pm 0.60$
    \\
    $\sigma_8$&$0.947^{+0.11}_{-0.045}$&$0.813^{+0.010}_{-0.012}$&$0.791^{+0.016}_{-0.021}$&$0.8120\pm 0.0091$
    \\
    $\Omega_\mathrm{m}$&$0.214^{+0.015}_{-0.072}$&$0.317^{+0.012}_{-0.0098}$&$0.339^{+0.025}_{-0.021}$&$0.3099\pm 0.0057$
    \\
    $S_8$&$0.781^{+0.021}_{-0.045}$&$0.835\pm 0.012$&$0.840^{+0.015}_{-0.013}$&$0.8253\pm 0.0097$
    \\
    \hline
    $\Delta\chi^2$&$-0.76$&$1.12$&$0.74$&$1.08$\\
    $\Delta\ln Z$&$0.11$&$-0.33$&$0.64$&$-0.21$ \\
    \hline\hline
    \end{tabular}
    \caption{We present the 68\% CL constraints on the cosmological parameters of the generalized DE scenario using various astronomical probes, including Planck 2018 CMB and other non-CMB datasets summarized in Section~\ref{sec:data}. Note that $\Delta\chi^2 > 0$ indicates a preference for our model over the CPL parametrization, while $\Delta\ln{Z} < 0$ implies that our model is also favored over CPL in terms of Bayesian evidence.}
    \label{tab:planck}
\end{table*}

\begin{table*}%[h]
    \centering
    \begin{tabular}{ccccccccccc}
    \hline 
    Parameters &AW-L & AW-LS & AW-LB & AW-LBS\\
    \hline\hline
    $100\theta_\mathrm{MC}$&$1.04089\pm 0.00029$&$1.04083\pm0.00028$&$1.04093\pm0.00028$&$1.04097\pm 0.00028$ 
    \\
    $\Omega_\mathrm{b} h^2$&$0.02265\pm0.00012$&$0.02263\pm 0.00012$&$0.02264\pm 0.00012$&$0.02265\pm 0.00012$
    \\
    $\Omega_\mathrm{c} h^2$&$0.1193^{+0.0013}_{-0.0015}$&$0.1201\pm 0.0012$&$0.11910\pm 0.00096$&$0.11865\pm 0.00096$
    \\
    $n_\mathrm{s}$&$0.9700\pm{0.0044}$&$0.9689\pm0.0041$&$0.9709\pm 0.0039$&$0.9719\pm 0.0039$
    \\
    $\tau_\mathrm{reio}$&$0.0561\pm0.0058$&$0.0568\pm 0.0056$&$0.0580\pm 0.0056$&$0.0595\pm 0.0056$
    \\
    $\log(10^{10} A_\mathrm{s})$&$3.048\pm0.011$&$3.052\pm 0.010$&$3.052\pm 0.010$&$3.055\pm 0.010$
    \\
    \hline
    $w_0$&$-1.04^{+0.36}_{-0.70}$&$-0.893\pm0.094$&$-0.57^{+0.27}_{-0.21}$&$-0.838^{+0.064}_{-0.084}$
    \\
    $w_\beta$&$<-0.385$&$-0.9^{+1.3}_{-1.0}$&$< -1.19$&$-1.04^{+1.0}_{-0.40}$
    \\
    $\beta$&---&$>6.42$&$2.5^{+1.0}_{-2.7}$&$3.7^{+3.4}_{-5.9}$
    \\
    \hline
    $H_0$&$77\pm10$&$67.07^{+0.85}_{-1.1}$&$65.3^{+1.9}_{-2.6}$&$67.76\pm 0.61$
    \\
    $\sigma_8$&$0.892^{+0.13}_{-0.087}$&$0.8119^{+0.0094}_{-0.011}$&$0.793^{+0.017}_{-0.021}$&$0.8117\pm 0.0087$
    \\
    $\Omega_\mathrm{m}$&$0.260^{+0.035}_{-0.12}$&$0.319^{+0.011}_{-0.0096}$&$0.335^{+0.025}_{-0.022}$&$0.3092\pm 0.0058$
    \\
    $S_8$&$0.803^{+0.037}_{-0.054}$&$0.837\pm0.011$&$0.837^{+0.015}_{-0.013}$&$0.8240\pm 0.0093$
    \\
    \hline
    $\Delta\chi^2$&$-0.08$&$-0.36$&$-0.22$&$0.37$\\
    $\Delta\ln Z$&$0.45$&$-0.41$&$0.56$&$-0.12$ \\
    \hline\hline
    \end{tabular}
    \caption{We present the 68\% CL constraints on the cosmological parameters of the generalized DE scenario using various astronomical probes, including the ACT DR6+WMAP CMB dataset and other non-CMB datasets summarized in Section~\ref{sec:data}. Note that $\Delta\chi^2 > 0$ indicates that our model is favored over the CPL parametrization, while $\Delta\ln{Z} < 0$ implies a preference for our model over CPL in terms of Bayesian evidence.  }
    \label{tab:act}
\end{table*}
\begin{figure*}
        \includegraphics[width=0.8\textwidth]{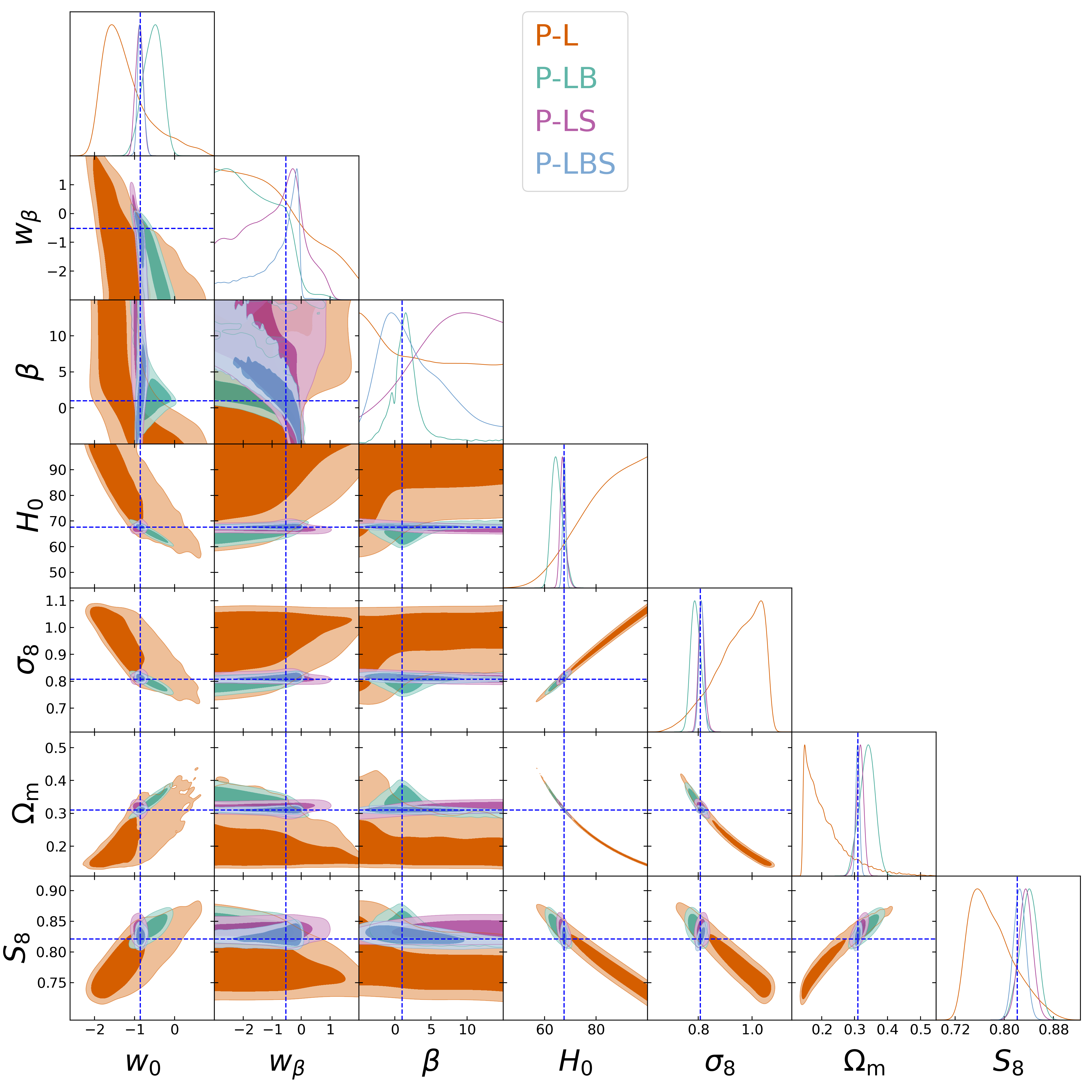}
        \caption{Constraints from dataset combinations including Planck 2018 CMB. The dotted line represents the best-fit values for the CPL parameterization with the P-LBS dataset combination.}
        \label{fig:triangle-planck}
        \end{figure*}
    \begin{figure*}
    \includegraphics[width=0.8\textwidth]{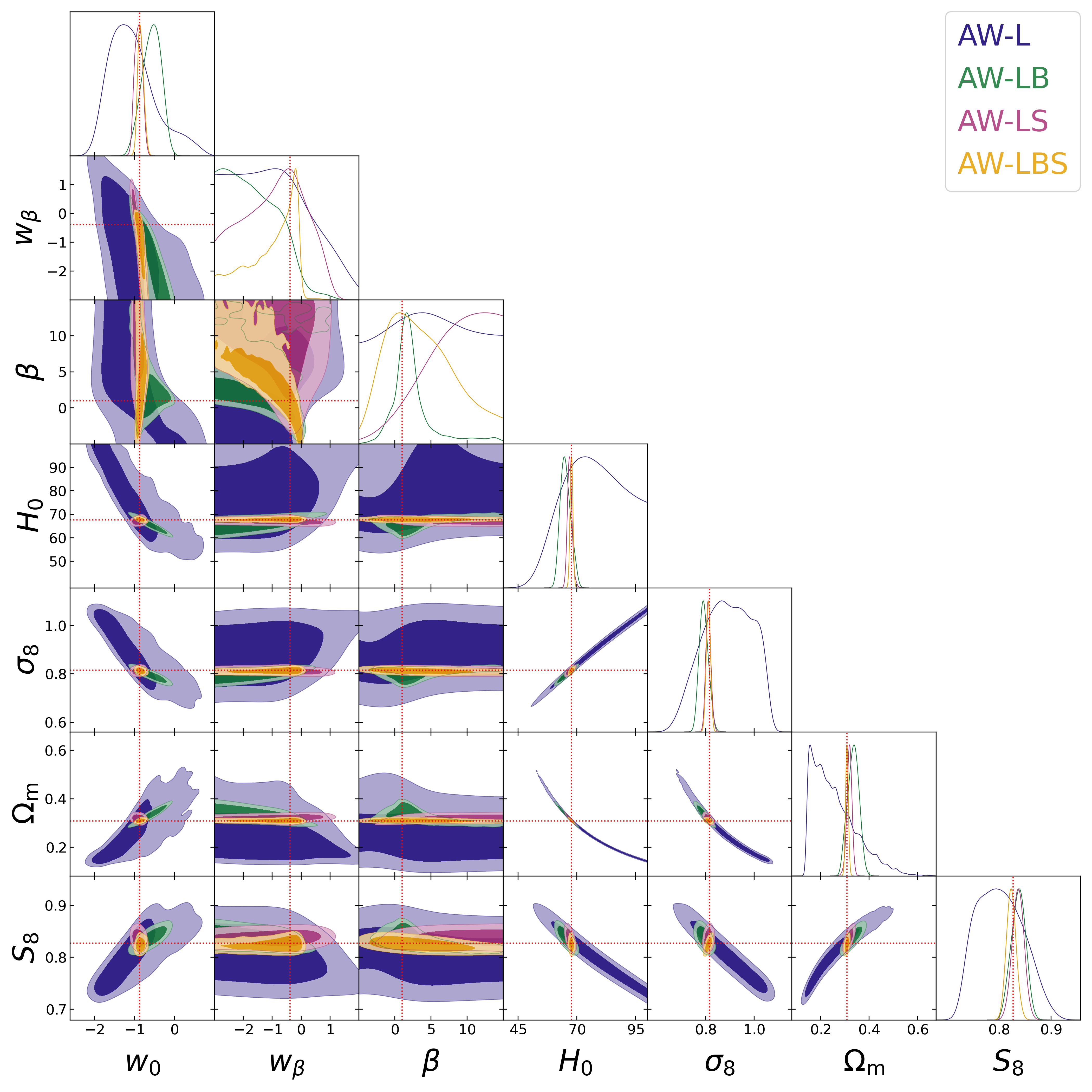}
    \caption{Constraints from dataset combinations including the ACT DR6 and WMAP CMB datasets. The dotted line represents the best-fit values for the CPL parameterization with the AW-LBS dataset combination.}
    \label{fig:triangle-act}
\end{figure*}
\begin{figure*}
    \centering
    \includegraphics[width=0.8\textwidth]{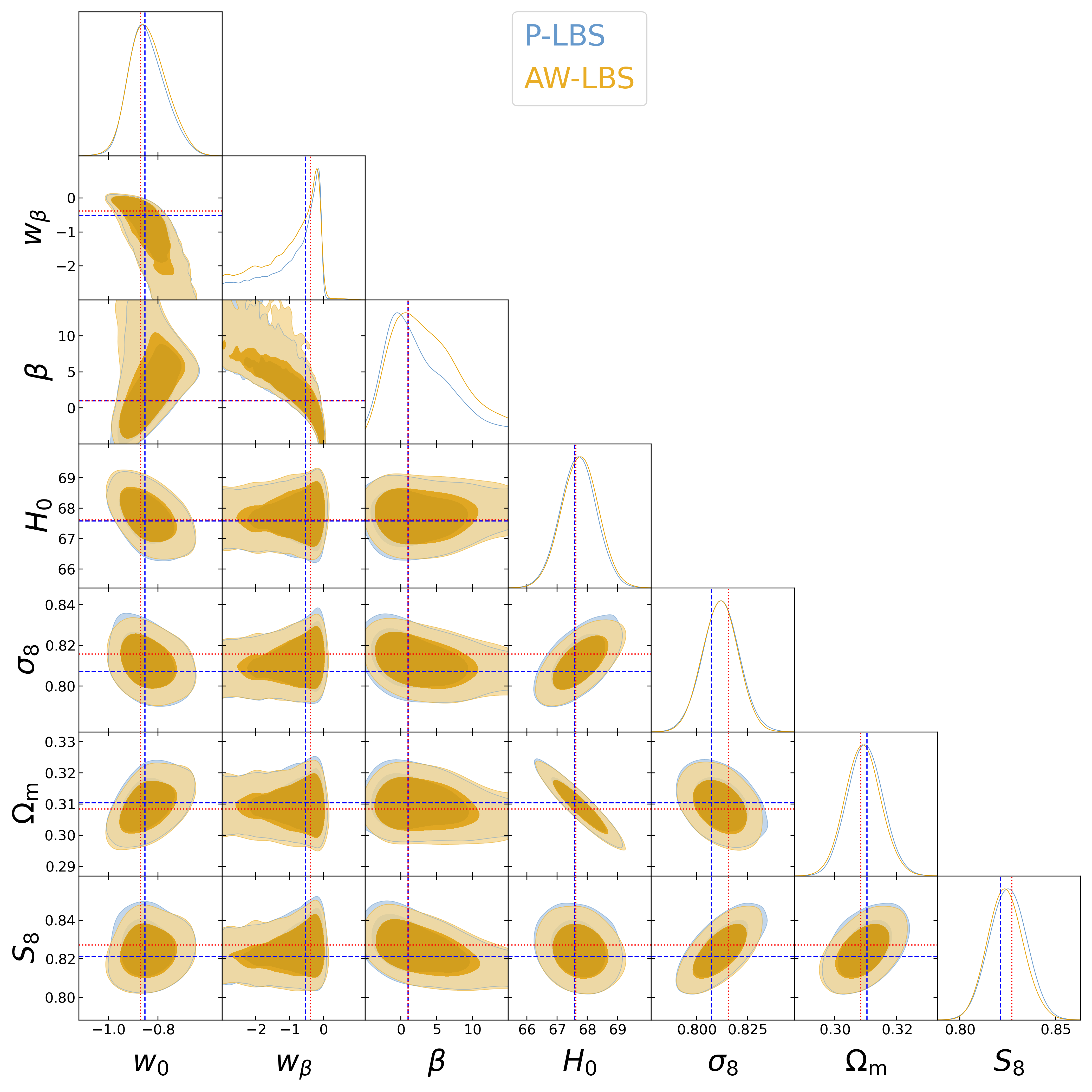}
\caption{One-dimensional marginalized posterior distributions and two-dimensional joint contours at 68\% and 95\% CL for the generalized DE scenario, obtained using the Planck CMB and ACT+WMAP CMB datasets, each combined with DESI and Pantheon+.}
    \label{fig:full}
\end{figure*}
\begin{figure*}
    \centering
    \subfloat[posterior plot of the equation of state $w(z)$.\label{fig:eos}]{%
        \includegraphics[width=0.49\textwidth]{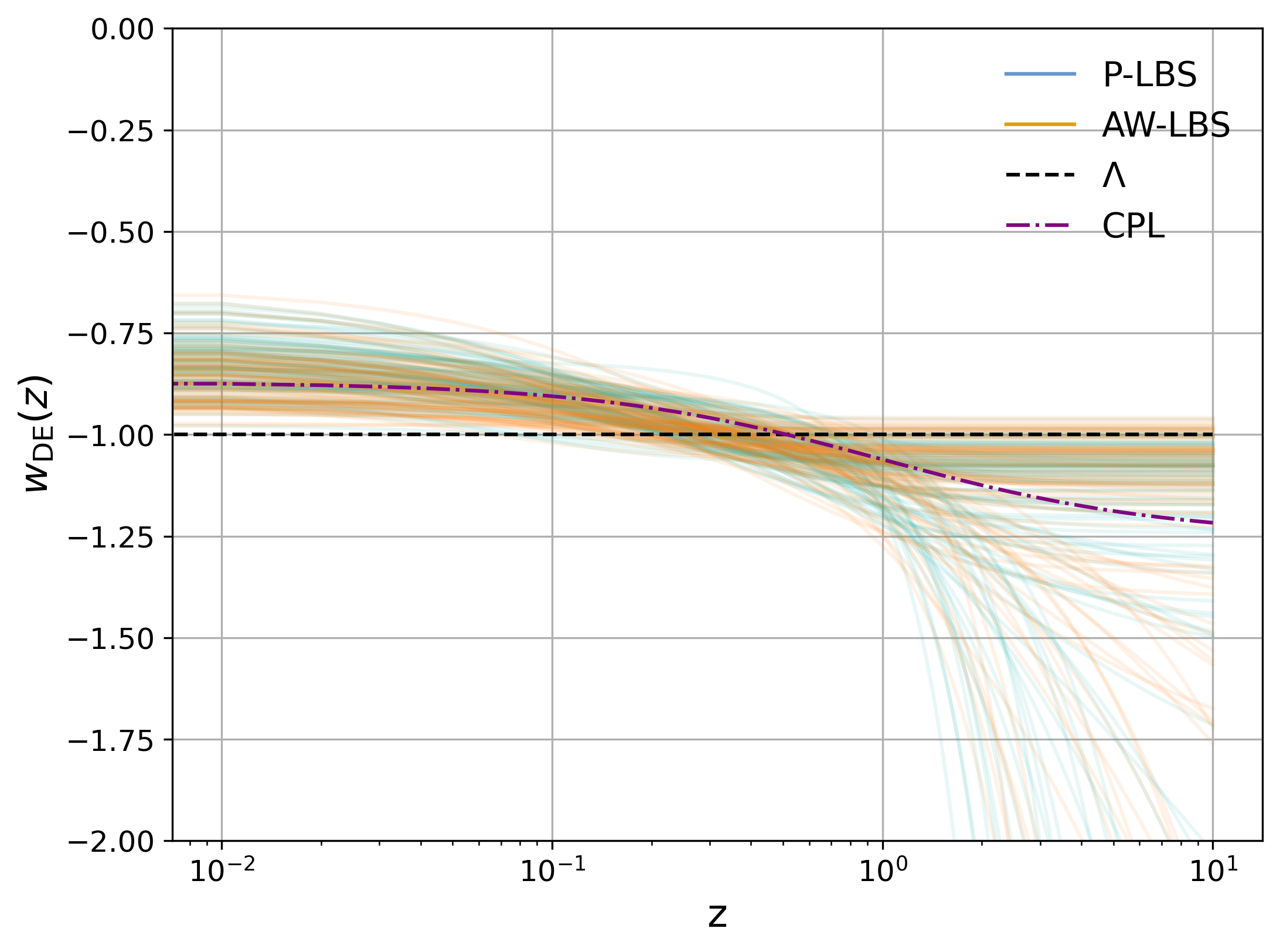}
    }
    \hfill
    \subfloat[posterior plot of the DE density evolution with respects to the present day $\rho_\mathrm{DE}(z)/\rho_{\mathrm{DE},0}$.\label{fig:rho}]{%
    \includegraphics[width=0.48\textwidth]{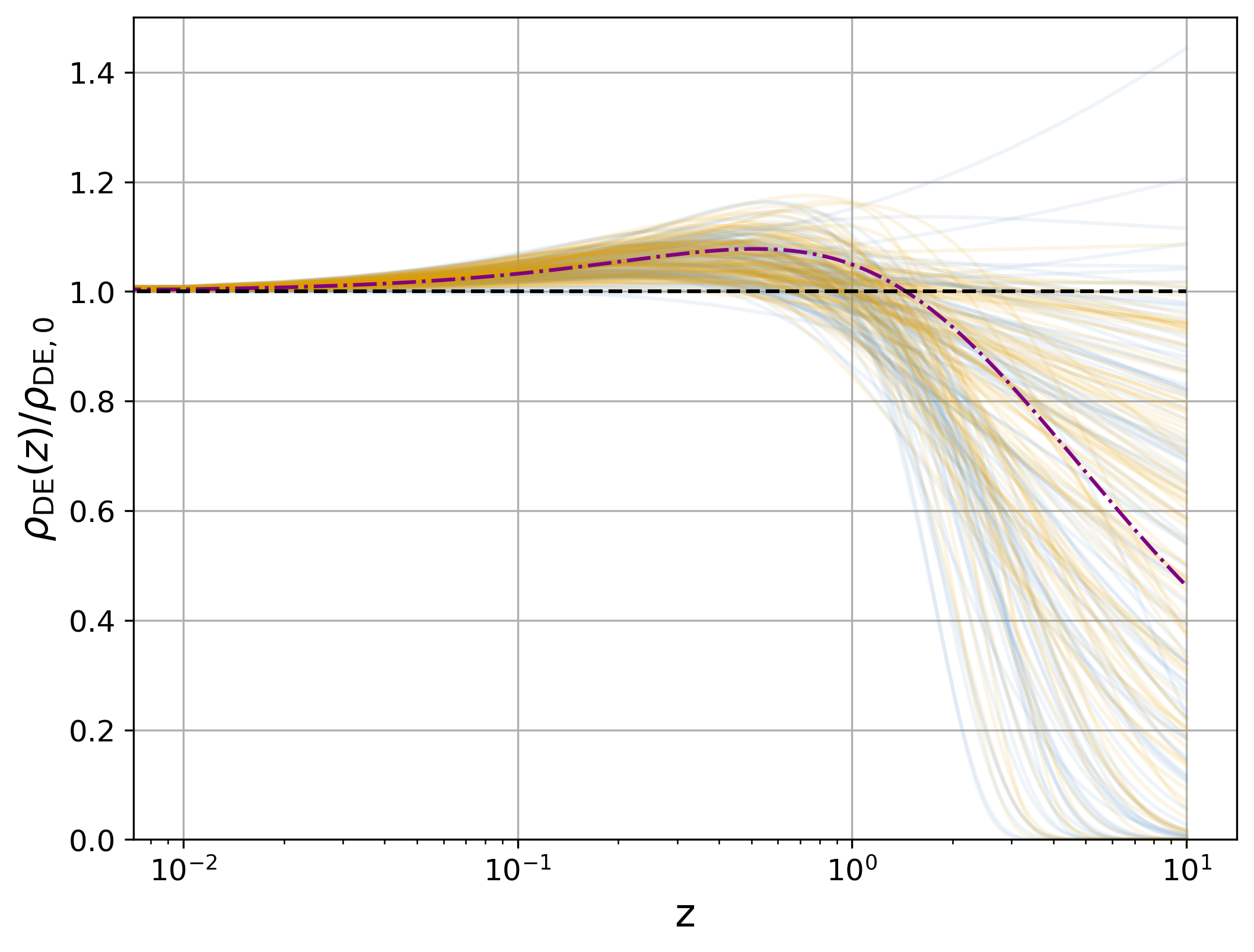}
    }
    \caption{Functional posterior plot using the full combination of datasets of CMB, CMB lensing, DESI Year 2 and Pantheon+. The CPL line represents the best-fit with the AW-LBS dataset.}
    \label{fig:eos-energy-density}
\end{figure*}

\section{Results: Observational constraints}
\label{sec-results}

In this section, we present the observational constraints on various free and derived parameters of the present DE model, considering several cosmological datasets described in Section~\ref{sec:data}. Additionally, we examine the statistical preference of the DE model with respect to the eight-parameter CPL model through both the $\Delta \chi^2$ statistic, defined as $\Delta \chi^2 = \chi^2_\mathrm{CPL} - \chi^2_\mathrm{model}$, and the Bayesian evidence (BE) comparison, characterized by the Bayes factor $\Delta\ln{Z} = \ln{Z_\mathrm{CPL}} - \ln{Z_\mathrm{model}}$. Accordingly, $\Delta \chi^2 > 0$ indicates that our model is favored over the CPL parametrization, while $\Delta \ln{Z} < 0$ implies that our model is preferred over CPL in terms of Bayesian evidence.

We choose CPL instead of $\Lambda$CDM as the reference model because our primary interest is to assess whether the present dynamical DE model can significantly improve the fit compared to CPL, particularly in light of recent results from the DESI collaboration, which suggest a preference for dynamical DE.  Due to this choice of model comparison, we would expect $\Delta\chi^2\geq0$ as the CPL parameterization is nested within our more general parameterization, while $\Delta\ln Z$ requires that the extended model demonstrate sufficient improvement over the simpler model.
The summary of the constraints is presented in Tables~\ref{tab:planck} and~\ref{tab:act}, and in Figs.~\ref{fig:triangle-planck}, \ref{fig:triangle-act}, \ref{fig:full}, and~\ref{fig:eos-energy-density}.

We start with the constraints presented in Table~\ref{tab:planck} for CMB data from the Planck 2018 dataset, combined with CMB lensing measurements from the ACT DR6 lensing likelihood and the Planck PR4 lensing data, as well as their combinations with other likelihoods, such as S and B, as labeled in Section~\ref{sec:data}.  And Fig. \ref{fig:triangle-planck} displays the correlations with some of the free and derived parameters.
We focus on the DE parameters $w_{0}$, $w_{\beta}$, and $\beta$, which collectively describe the nature of the DE parametrization, along with the derived parameters $H_0$, $\Omega_{\rm m}$, and $S_8$. 
We find that the P-L dataset is unable to constrain $w_{\beta}$ and $\beta$, although $w_0$ shows a tendency to lie in the phantom regime—an observation already noted in the literature~\cite{DiValentino:2016hlg}. However, within the 95\% CL, $w_0$ remains consistent with $-1$. Since both $w_{\beta}$ and $\beta$ remain unconstrained in this case, no definitive conclusion about the nature of DE can be drawn from CMB data alone.
Due to the phantom-like tendency of $w_0$, the expansion rate of the universe increases, resulting in a higher lower bound for $H_0$. This, in turn, leads—through the geometric degeneracy between $H_0$ and $\Omega_{\rm m}$—to a lower inferred value of $\Omega_{\rm m}$, with $\Omega_{\rm m} = 0.214^{+0.14}_{-0.075}$ at 95\% CL (for P-L). At the same time, we observe a slightly lower value of $S_8$, namely $S_8 = 0.781^{+0.021}_{-0.045}$ at 68\% CL, which is lower than the $\Lambda$CDM-based Planck 2018 estimate~\cite{Planck:2018vyg}.

When the Pantheon+ dataset is added to P-L (labeled as P-LS in Table~\ref{tab:planck}), we find that $w_0$ shifts to the quintessence regime. However, within the 95\% CL, the value $w_0 = -1$ is still allowed. On the other hand, $\beta$ remains unconstrained, while $w_\beta$ now attains a lower bound. Therefore, no conclusive evidence for the dynamical nature of the EoS can be inferred.
The constraints on the derived parameters ($H_0$, $\Omega_{\rm m}$, and $S_8$) are nearly consistent with the $\Lambda$CDM-based Planck 2018 estimates~\cite{Planck:2018vyg}, indicating that this combination does not strongly depart from the standard cosmological model.

The inclusion of BAO with P-L proves particularly interesting, as this is the first case in which $\beta$ becomes constrained: $\beta = 2.0^{+1.2}_{-2.6}$ at 68\% CL (P-LB). However, $w_\beta$ still has only a lower bound, and $w_0$ remains in the quintessence regime, while marginally allowing $w_0 = -1$ ($w_0 = -0.54^{+0.43}_{-0.46}$ at 95\% CL).
In this case, $\beta$ spans a broad range, $(-0.6, 3.2)$, covering both $\beta = 1$ and the limit $\beta \rightarrow 0$, indicating that the P-LB dataset is compatible with various dynamical DE scenarios. On the other hand, the linear parametrization with $\beta = -1$ is ruled out, as is $\Lambda$CDM, corresponding to the limit $\beta \rightarrow \infty$.
Nevertheless, since $w_\beta$ remains bounded only from below, no definitive conclusion regarding the dynamical nature of $w_{\rm DE}$ can yet be drawn.

When both the BAO and Pantheon+ datasets are included with P-L, the parameters $w_{\beta}$ and $\beta$ become simultaneously constrained. This combination provides the most stringent constraints on the model  (see Fig. \ref{fig:full}). In this case, we observe that $w_0$ remains strictly in the quintessence regime at slightly more than 95\% CL, with $w_0 = -0.84^{+0.15}_{-0.13}$ at 95\% CL. 
The key parameter $\beta$ allows for a wide range of possibilities due to its large uncertainties ($\beta = 2.8^{+2.8}_{-5.9}$ at 95\% CL), and notably, the condition $\beta\rightarrow 0$ is also included within this range. 
The parameter $w_{\beta}$ marginally allows a null value ($w_{\beta} = -0.94^{+0.94}_{-0.31}$ at 95\% CL), indicating that $w$CDM is marginally permitted by this dataset at 95\% CL. 
The constraints on the usual derived parameters ($H_0$, $\Omega_{\rm m}$, and $S_8$) are nearly identical to those found within the $\Lambda$CDM framework as inferred from Planck~\cite{Planck:2018vyg}.

Having summarized the constraints from different datasets, we now focus on the statistical preference of the DE model with respect to the CPL parametrization.
Following the sign convention for $\Delta \chi^2$ and $\Delta\ln{Z}$, we find that for the P-L dataset, both indicators slightly disfavor the generalized DE model compared to CPL, with $\Delta\ln{Z} > 0$.

In contrast, the P-LS ($\Delta \chi^2=1.12$, $\Delta\ln{Z} = -0.33$) and P-LBS ($\Delta \chi^2=1.08$, $\Delta\ln{Z} = -0.21$) datasets favor the present DE model, as they yield $\Delta \chi^2 > 0$ and $\Delta\ln{Z} < 0$.
For the P-LB dataset, the model is preferred according to $\Delta \chi^2$, but disfavored according to $\Delta\ln{Z}$. 
Interestingly, the P-LBS combination provides the most stringent constraints on the model and shows a consistent statistical preference for the DE parametrization over CPL. 
However, we note that in all cases, the absolute values of $\Delta\ln{Z}$ remain below 1, which, according to the revised Jeffreys scale, corresponds to inconclusive evidence. Thus, while trends are suggestive, the Bayesian model comparison does not yield statistically decisive support for either model.

We now focus on the constraints obtained using the alternative CMB dataset (ACT DR6 + WMAP) and its combination with similar low-redshift datasets. In Table~\ref{tab:act}, we present the constraints on the free and derived parameters of the DE model for this case and Fig.~\ref{fig:triangle-act} presents the relevant graphical descriptions.
Considering only the base dataset (i.e. AW-L) in this series of analyses, we observe that $w_0$ shows a phantom tendency. However, within the 68\% CL, values in the quintessence regime are also allowed. On the other hand, $w_\beta$ is constrained from above, while $\beta$ remains unconstrained. 
This represents a partial improvement over the P-L dataset, where both $w_\beta$ and $\beta$ were unconstrained (see the second column of Table~\ref{tab:planck}). The derived parameters in this case follow a similar, though not identical, pattern to that observed with the P-L dataset: we see an increase in $H_0$, accompanied by a corresponding decrease in $\Omega_m$ and $S_8$.

When Pantheon+ is combined with AW-L (i.e. the combined dataset labeled as AW-LS), we find that $w_0$ lies in the quintessence regime, although not strictly, as $w_0 = -1$ is still allowed within the 95\% CL. On the other hand, $w_\beta$ is now constrained, with $w_\beta = -0.9^{+1.6}_{-2.1}$ at 95\% CL for AW-LS. This is an improvement over the P-LS case, which only provided a lower bound, $w_\beta < 0.521$ at 95\% CL.
Additionally, the $\beta$ parameter now has a lower bound, $\beta > 0.309$ at 95\% CL, whereas it remained unconstrained for P-LS. These results indicate that the constraining power of the AW-LS combination is slightly improved compared to P-LS.

The inclusion of DESI BAO with AW-L allows for the constraint of the parameter $\beta$, yielding $\beta = 2.5^{+1.0}_{-2.7}$ at 68\% CL. However, $w_{\beta}$ is found to have only an upper limit, and the present-day value of the DE equation of state, $w_0$, lies in the quintessence regime at more than 95\% CL.
Given the large uncertainties, $\beta$ spans a wide interval [$\beta \in (-0.2, 3.5)$], and therefore, from a statistical perspective, various possibilities remain viable—including $\beta = 1$ (corresponding to the CPL parametrization) and the limit $\beta \rightarrow 0$ (logarithmic parametrization). On the other hand, the linear parametrization with $\beta = -1$ is ruled out, as is $\Lambda$CDM, which corresponds to the limit $\beta \rightarrow \infty$.
Since $w_{\beta}$ remains unconstrained in this case, any conclusive statement about the dynamical nature of $w_{\rm DE}$ would be premature.  

When both BAO and Pantheon+ are combined with AW-L, we find that the parameters $w_{\beta}$ and $\beta$ are simultaneously constrained—similar to what was observed earlier with the P-LBS dataset  (see Fig. \ref{fig:full}). This clearly indicates that, among all the data combinations considered, AW-LBS provides the most stringent constraints on this DE scenario, as in other cases only one of the two parameters is constrained.
We observe that the mean value of $w_0$ shifts closer to $-1$, yet it still lies in the quintessence regime at more than 95\% CL. Evidence for $w_{\beta} \neq 0$ is found at slightly more than 68\% CL, providing a hint of dynamical dark energy.
The parameter $\beta$ spans a wide interval due to large uncertainties ($\beta = 3.7^{+3.4}_{-5.9}$ at 68\% CL), and values such as $\beta = 1$, $\beta = -1$, and $\beta \rightarrow 0$ are all statistically allowed in this case.

Now, regarding the statistical preference of this DE model, our observations are as follows. For both the AW-L and AW-LB datasets, we find that $\Delta\chi^2<0$ and $\Delta\ln{Z} > 0$, indicating that the DE model is disfavored compared to the CPL parametrization.
For the AW-LS and AW-LBS datasets, instead, the DE model is preferred over CPL according to the BE. As noted earlier, the AW-LBS dataset also provides the most stringent constraints on the model parameters, which makes this result particularly noteworthy.
Nonetheless, we emphasize that in all cases, the absolute value of $\Delta\ln{Z}$ remains below 1, which corresponds to an \textit{inconclusive} level of statistical significance according to the revised Jeffreys scale. Therefore, although the trends are indicative, they do not constitute decisive Bayesian evidence in favor of either model.

Finally, we depict the evolution of the DE EoS $w_{\rm DE}(z)$ and its energy density $\rho_{\rm DE}(z)$  using the posteriors of the functions given in Eqs.~(\ref{general-eos}) and~(\ref{rho-DE}), respectively, plotted with the functional posterior tool \texttt{fgivenx}~\cite{fgivenx}.
Figure~\ref{fig:eos} shows that, regardless of the CMB dataset used, $w_{\rm DE}(z)$ exhibits a transition (crossing of the phantom divide line) from an early-time phantom behavior to a present-day quintessence-like regime—consistent with the trend also reported by DESI DR2~\cite{DESI:2025zgx} (see also~\cite{Ozulker:2025ehg}). 
From Fig.~\ref{fig:rho}, we observe that the dynamical evolution of $\rho_{\rm DE}$ was more pronounced in the past, whereas in recent times it tends to approach a constant value. This behavior suggests consistency with a cosmological constant at late times, as indicated by the full combined datasets.

\section{Summary and Conclusions}
\label{sec-summary}

Following the recent BAO measurements from DESI, DE models have come to the forefront of modern cosmology, with the question of whether DE is time-dependent becoming particularly significant. DESI has reported evidence for dynamical DE under the assumption of the CPL parametrization~\cite{DESI:2024mwx,DESI:2025zgx}, which has subsequently been applied to other DE parametrizations, yielding similar indications of dynamical behavior~\cite{Giare:2024gpk,Wolf:2025jlc,Cheng:2025lod}.
In light of this growing interest, in the present work we have considered a generalized DE parametrization characterized by three free parameters: $w_0$ (the present-day value of the DE equation of state), $w_{\beta}$ (capturing its dynamical nature at late times), and $\beta$ (which generates a variety of DE behaviors). For different values of $\beta$, this framework recovers a wide range of known DE parametrizations—such as CPL for $\beta = 1$, linear for $\beta = -1$, and logarithmic for $\beta \rightarrow 0$, as well as new, unexplored forms.
We have constrained this DE scenario using a comprehensive set of cosmological probes, including CMB data from Planck 2018; CMB from ACT DR6 combined with WMAP data and a Gaussian prior on the optical depth $\tau_{\rm reio} = 0.0566 \pm 0.0058$; CMB lensing from ACT DR6 and Planck PR4 (NPIPE); BAO measurements from DESI DR2; and Type Ia supernovae from the Pantheon+ compilation. The results are summarized in Tables~\ref{tab:planck} and~\ref{tab:act}, where we present the 68\% CL constraints on the model parameters.
Our key findings are the following:

\begin{itemize}
    \item Among all dataset combinations explored in this work, only two—P-LBS and AW-LBS—are able to simultaneously constrain all three DE parameters and are highly consistent with one another (see Fig.~\ref{fig:full}). The resulting constraints are: $w_0 = -0.842^{+0.060}_{-0.083}$, $w_{\beta} = -0.94^{+0.94}_{-0.31}$, and $\beta = 2.8^{+2.8}_{-5.9}$ for P-LBS; and $w_0 = -0.838^{+0.064}_{-0.084}$, $w_{\beta} = -1.04^{+1.0}_{-0.40}$, and $\beta = 3.7^{+3.4}_{-5.9}$ for AW-LBS (all at 68\% CL). In both cases, $w_0$ remains in the quintessence regime at more than 95\% CL—a result consistent with recent DESI DR2 findings~\cite{DESI:2025zgx}. There is also marginal evidence for $w_{\beta} \neq 0$ at approximately 95\% CL, indicating a possible dynamical nature of DE. However, the parameter $\beta$ remains poorly constrained and spans a wide range, including multiple known DE parametrizations, preventing any conclusive interpretation on a preferred DE model at this stage.

    \item Except for the P-L and AW-L datasets, the derived parameters $H_0$, $\Omega_{\rm m}$, and $S_8$ show values consistent with the $\Lambda$CDM predictions reported by Planck~\cite{Planck:2018vyg}.

    \item In agreement with recent DESI measurements~\cite{DESI:2025zgx} (see also~\cite{Ozulker:2025ehg}), we observe a transition of $w_{\rm DE}(a)$ from an early phantom regime to a more recent quintessence-like behavior (see Fig.~\ref{fig:eos}).

    \item Based on the $\Delta \chi^2$ statistic and Bayesian evidence (quantified via $\ln Z$), we find that for both full combined datasets (P-LBS and AW-LBS), our present dynamical DE model is favored over the CPL parametrization. However, we note that in both cases, the absolute value of $\Delta \ln Z$ remains below 1, which—according to the revised Jeffreys scale—corresponds to an \textit{inconclusive} level of statistical significance.
\end{itemize}

The results on the generalized DE parametrization summarized above are important for two specific reasons. First, to the best of our knowledge, this is the first time the model has been analyzed while accounting for its evolution at the perturbation level, which is essential for a complete cosmological analysis. Second, despite having three free parameters, both the $\chi^2$ analysis and the Bayesian evidence consistently favor this model over the widely studied CPL parametrization.
With the recent release of DESI DR2 and several upcoming data releases expected in the near future, it is clear that DE phenomenology will remain a central topic in modern cosmology. Last but not least, since the present DE model is flexible enough to encompass a broad class of DE parametrizations, many intriguing results related to this framework are likely still to come.

\section*{Acknowledgments}
We thank the referee for some useful comments and suggestions which helped us to improve the manuscript.
DHL is supported by an EPSRC studentship. WY has been  supported by the National Natural Science Foundation of China under Grants No. 12175096, and Liaoning Revitalization Talents Program under Grant no. XLYC1907098. EDV acknowledges support from the Royal Society through a Royal Society Dorothy Hodgkin Research Fellowship. SP acknowledges the financial support from the Department of Science and Technology (DST), Govt. of India under the Scheme ``Fund for Improvement of S\&T Infrastructure (FIST)'' (File No. SR/FST/MS-I/2019/41).  CvdB is supported (in part) by the Lancaster–Sheffield Consortium for Fundamental Physics under STFC grant: ST/X000621/1.
We acknowledge the IT Services at The University of Sheffield for the provision of services for High Performance Computing. 
This article is based upon work from the COST Action CA21136 - ``Addressing observational tensions in cosmology with systematics and fundamental physics (CosmoVerse)'', supported by COST - ``European Cooperation in Science and Technology''.

%\clearpage
%---------------------------------------------------------------
\bibliography{biblio}
%---------------------------------------------------------------
\end{document}